# Chiral Light Emission from a Sphere Revealed by Nanoscale Relative Phase Mapping

*Taeko Matsukata[1,2], F. Javier García de Abajo[3,4], Takumi Sannomiya[1,5]*

**AUTHOR ADDRESSES**

[1] Department of Materials Science and Technology, Tokyo Institute of Technology, 4259 Nagatsuta Midoriku, Yokohama 226-8503, Japan

[2] RIKEN, 2-1 Hirosawa, Wako, Saitama 351-0198, Japan

[3] ICFO-Institut de Ciencies Fotoniques, The Barcelona Institute of Science and Technology, 08860 Castelldefels (Barcelona), Spain

[4] ICREA-Institució Catalana de Recerca i Estudis Avancats, Passeig Lluís Companys 23, 08010 Barcelona, Spain

[5] PRESTO, 4259 Nagatsuta Midoriku, Yokohama 226-8503, Japan

**ABSTRACT**

Circularly polarized light (CPL) is currently receiving much attention as a key ingredient for next-generation information technologies, such as quantum communication and encryption. CPL photon generation used in those applications is commonly realized by coupling achiral optical quantum emitters to chiral nanoantennas. Here, we explore a different strategy consisting in exciting a nanosphere --the ultimate symmetric structure-- to produce CPL emission along an arbitrary direction. Specifically, we demonstrate chiral emission from a silicon nanosphere induced by an electron beam based on two different strategies: dissolving the degeneracy of orthogonal dipole modes, and interference of electric and magnetic modes. We prove these concepts both theoretically and experimentally by visualizing the phase and polarization using a polarimetric four-dimensional cathodoluminescence method. Besides their fundamental interest, our results support the use of free-electron-induced light emission from spherically symmetric systems as a versatile platform for the generation of chiral light with on-demand control over the phase and degree of polarization.

**TOC FIGURE**

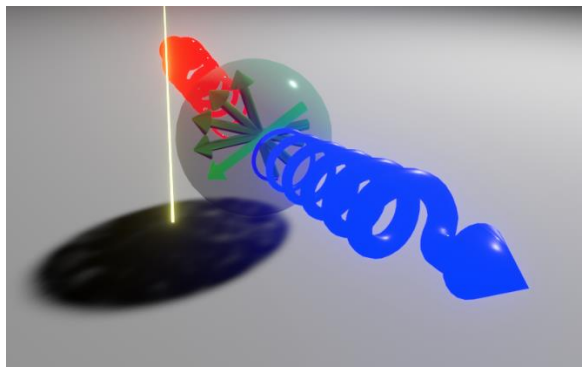

Circularly polarized light (CPL) has been identified as a key ingredient for next-generation information technologies, such as quantum communication[1, 2] and encryption.[3, 4] CPL, classified by its parity as having right- or left-handed circular polarization (RCP, LCP), can be utilized as a digital signal carrier.[5, 6] Compared to linearly polarized light, which can convey polarization information in an analog manner, CPL provides a more robust platform for information transfer that has been adopted even in biological systems.[7, 8] However, light sources used in quantum circuits, such as single-photon generators, consist of linear optical dipoles typically hosted on molecules, atomic defects, or few-nanometer quantum dots, which are not capable of encoding CPL information unless one involves momentum transfer to electron spins under high magnetic field and low temperature.[5] A routinely used way to efficiently generate CPL from atomic-scale emitters consists in coupling them to chiral nanoantennas.[9-11] This strategy allows using any light source without chirality to generate CPL. Various chiral optical nanoantennas have been proposed, which include three-dimensionally sculpted plasmonic nanostructures with chiral structural symmetry, typically shaped as helical structures[12] inspired by radio frequency antennas, or multiple nanoparticles stuck in layers to support collective geometrical chirality.[13-15] Instead of using such intrinsically chiral structures, chiral emission from two-dimensional achiral structures,[16-18] such as a rectangle and an ellipse, have recently been reported in which chiral emission is obtained by breaking mirror symmetry through external factors, such as localized excitation and detection. This is the so-called extrinsic chirality, which is associated with an externally broken symmetry.[10] Achiral nanoantennas present the advantages of being equally responsive to both CPL polarizations, yet offering some selectivity through the local field and providing an efficient solution as CPL light emitters and receivers.[19] Spherical nanoparticles, which constitute ultimate examples of symmetric structures, work as omnidirectional antennas, but of course without intrinsic chirality. Although a spherical structure has never been used as a CPL light source, chiral emission from a spherical nanoantenna can be obtained through extrinsic chirality supported by breaking the symmetry of the whole excitation and measurement system. In general, CPL generation can be ascribed to the interference between two orthogonal linear dipoles with a relative phase difference. In previous studies of rectangular or elliptical achiral nanoantennas,[16-18] two orthogonal linear dipoles with different resonance energies are utilized to produce a relative phase difference.[16-18] However, the dipole modes of a sphere are degenerate and the phase difference must be generated through a properly designed excitation.[20] Free electron beams provide a versatile way of exciting optical modes with high spatial resolution that has been extensively used to investigate the optical response of different structures from the resulting cathodoluminescence (CL) light emission.[21-25] This approach thus presents potential advantages for the on-demand generation of CPL, which have not so far been exploited.

Here, we propose two possible schemes to generate CPL from a sphere induced by free electron beams. The first approach consists in shifting the phase of orthogonal degenerate modes[26, 27] as schematically illustrated in Figure 1a and c. In a second approach, we rely on two modes with different resonance energies, and more precisely, the electric and magnetic dipoles supported by the dielectric nanosphere, which can generate CPL through mutual interference, as depicted in Figure 1b and d.

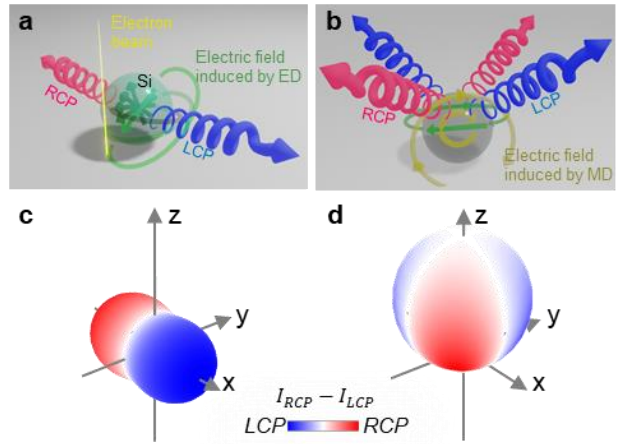

**Figure 1. Illustration of CPL emitted from a spherical dielectric particle.** Conceptual diagrams of CPL generated (a) by a perpendicular (along the *z* axis) electric dipole (*p*ED) and an in-plane (in the *x*-*y* plane) electric dipole (*i*ED) with different phases excited through a swift electron; and (b) by interfering *i*ED and *i*MD (in-plane magnetic dipole) modes. (c-d) CPL handedness projected on the radiation patterns of two interfering orthogonal linear dipoles of the same magnitude with different relative phases: (c) *i*ED + *p*ED with a phase shift of π; and (d) *i*ED + *i*MD with a phase shift of π/4.

Knowledge about the nanoscopic field distribution and the relative phase of excitation modes is essential to design achiral nanoantennas. In order to access the field distribution inside the structure, which is especially important for dielectric antennas, and detect chiral emission, scanning transmission electron microscopy (STEM) with CL[28] is an ideal choice because no other currently available nano-optical measurement techniques satisfy that requirement; in particular, scanning near-field optical microscopy (SNOM)[29] cannot probe the internal field, while electron energy loss spectroscopy (EELS) has no chiral detection capabilities. Indeed, we have previously shown that STEM-CL can visualize the interference of emitted light.[30, 31] In this work, we demonstrate chiral emission from a Si nanosphere using STEM-CL by visualizing the nanoscopic relative phase distribution of the chiral excitation state. In order to do so, we have developed a four-dimensional (4D) STEM-CL technique capable of rendering simultaneous angle- and energy-resolved field mapping with full polarimetry (Figure 2). We further validate the experimental results through rigorous CL calculations based on an analytical multipole decomposition (AMD) method.

**RESULTS & DISCUSSION**

**4D STEM-Cathodoluminescence**

Since the CL signal is proportional to the projection of the radiative component of the electromagnetic local density of state along the trajectory of the electron beam,[26, 32] the detected CL signal from a sphere corresponds to the superposition of multiple modes that can be selectively observed by resolving the detection angle.[33] The 4D STEM-CL setup that we use in this study allows imaging with simultaneous acquisition of the angular distribution and spectrum at each beam position (corresponding to 51 (angle) × 256 (energy) images), as described in Figure 2a. The CL signal collimated by a parabolic mirror accommodated in the STEM column is collected by the 2D CCD elements of a spectrometer. By inserting a slit mask in the light path, the radiation in the azimuthal angle is selected with $\varphi = 0°$, and is transferred through the optical system to the CCD with information on the $\theta$-angle dependence preserved along the vertical axis, while information on photon energy is obtained through horizontal dispersion using a grating. The resulting 2D angular-energy information is thus acquired while raster-scanning the electron beam in 2D space, leading to 4D (1D angle, 1D energy, and 2D space) CL measurements (see more details on our apparatus in Methods section). To observe CPL from a Si sphere by electron beam excitation we performed fully polarimetric imaging and obtained CL maps with six polarization configurations for the Stokes parameter calculation. The circular polarization state (RCP or LCP) is selected by a polarizer and a quarter-wave plate (QWP). The polarization angle $\zeta$ is defined from the vertical axis, as shown in Figure 2a. In what follows, *p*- and *s*-polarizations denote $\zeta = 0°$ and $\zeta = 90°$, respectively. The Stokes parameters ($S_{0-3}$) can be obtained from the intensities of the linearly polarized signal $I_\zeta$ (polarization angle $\zeta$) and circularly polarized signals $I_{\text{RCP}}$ and $I_{\text{LCP}}$ as $S_0 = I_{\text{non}} = I_s + I_p$, $S_1 = I_s - I_p = I_{\text{non}}\, p \cos 2\alpha \cos 2\eta$, $S_2 = I_{45°} - I_{-45°} = I_{\text{non}} p \sin 2\alpha \cos 2\eta = 2 I_s I_p \cos \delta$, $S_3 = I_{\text{RCP}} - I_{\text{LCP}} = I_{\text{non}}\, p \sin 2\eta = 2 I_s I_p \sin \delta$, and $p = \sqrt{S_1^2 + S_2^2 + S_3^2}/S_0$. Here, we introduce the ellipticity angle $\eta$ and orientation angle $\alpha$ as defined in Figure 2b. The ellipticity angle $\eta$ ranges from $-\pi/4$ to $\pi/4$ and is defined such that positive and negative values correspond to RCP and LCP, respectively. We also investigate the phase difference $\delta$ between the horizontal and vertical emitted electric field, which can be calculated as $\delta = \arg(S_3/S_2)$. Figure 2c shows the experimentally obtained spectrum of a 250 nm Si sphere without polarization. The CL signal is integrated over the detection angle $\theta = 0 - 180°$ and spatially integrated over the Si sphere to include all excited modes. The observed resonance peaks are identified by comparison with AMD theory (see details in Methods and SI) and are labeled as magnetic dipole (MD), magnetic quadrupole (MQ), electric dipole (ED), and electric quadrupole (EQ). Dielectric particles can accommodate multiple field nodes in the radial direction inside the particle with the same symmetry in the angular direction, which we refer to as high-radial-order modes.[27] We indicate radial orders by superscripts (*e.g.*, $ED^2$ for the second-order electric dipole).

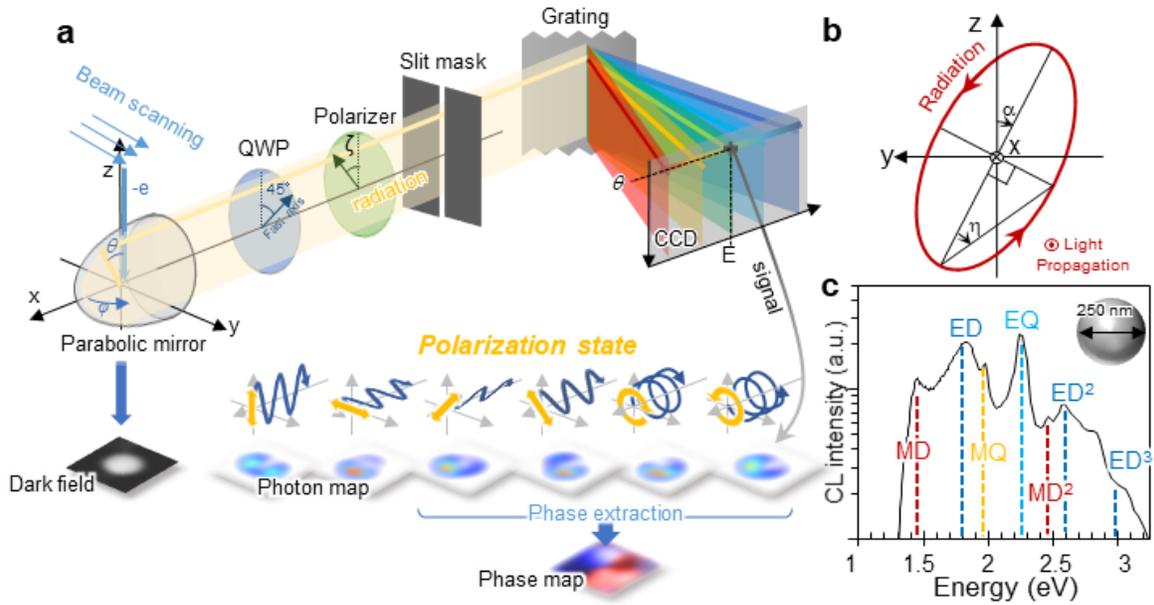

**Figure 2.** (a) Schematic illustration of the 4D STEM-CL setup and diagrams of phase extraction. (b) Definition of the ellipticity angle *η* and the orientation angle *α*. (c) Experimentally observed CL spectrum from a 250 nm Si sphere without polarization selection, obtained by integrating the signal over the emission angle range *θ* = 0 - 180° and averaging over all electron beam positions on the sample. Mode types in the spectral features are indicated by labels: MD (magnetic dipole), ED (electric dipole), MQ (magnetic quadrupole), and EQ (electric quadrupole). The superscript denotes the radial order.

**CPL Emission and Phase Mapping of Interfering Degenerate Electric Dipoles**

The 4D STEM-CL enables visualizing photon map stacks fully resolved in angle and energy (see examples in Figure S4-6). In this section, we focus on CPL generation from EDs. We denote the in-plane ED with orbital momentum numbers (*l*, *m*) = (1, ±1) as *i*ED (see a more precise definition of ($l$, $m$) in SI); these modes present polarization lying on the *x-y* plane. The perpendicular ED characterized by ($l$, $m$) = (1, 0) and denoted *p*ED presents polarization along the *z* axis. With the electron beam running along the *z* axis, *i*ED and *p*ED modes are excited with phase mismatch even though they are degenerate;[26] we thereby generate CPL emission due to the interference of two orthogonal degenerate EDs, as described in Figure 1. Figure 3 shows the fully polarimetric CL maps acquired from a 250 nm Si nanosphere for selected photon energies of 1.8, 2.6, and 2.96 eV, corresponding to the resonances $ED^1$, $ED^2$, and $ED^3$, respectively (see Figure 2c and Figure S2a). We chose a detection angle $\theta = 135°$ (schematically shown in Figure 3b) as a compromise that fulfills both a sufficient signal intensity and interference of *i*ED and *p*ED. We further describe below the advantage of this detection angle to map the phase. At the photon energy of 1.8 eV for the $ED^1$ mode, the CL maps obtained by selecting linear polarization show clear dipolar hot spots along the polarization direction. The $S_3$ plot, which intuitively indicates the handedness of CPL, shows the generation of CPL, where the parity of CPL is inverted with respect to the *x-z* plane. This map proves the concept of CPL generation from a sphere controlled by the electron beam as illustrated in Figure 1a. The presented CPL generation implies an effective phase difference of degenerate *p*ED and *i*ED modes excited by a swift electron, which we further discuss below by performing phase mapping. For the higher-radial-order EDs at photon energies of 2.6 eV ($ED^2$) and 2.96 eV ($ED^3$), we observe the similar features, except that they show double- and triple-layered dipole patterns with parity inversion also along the radial direction.

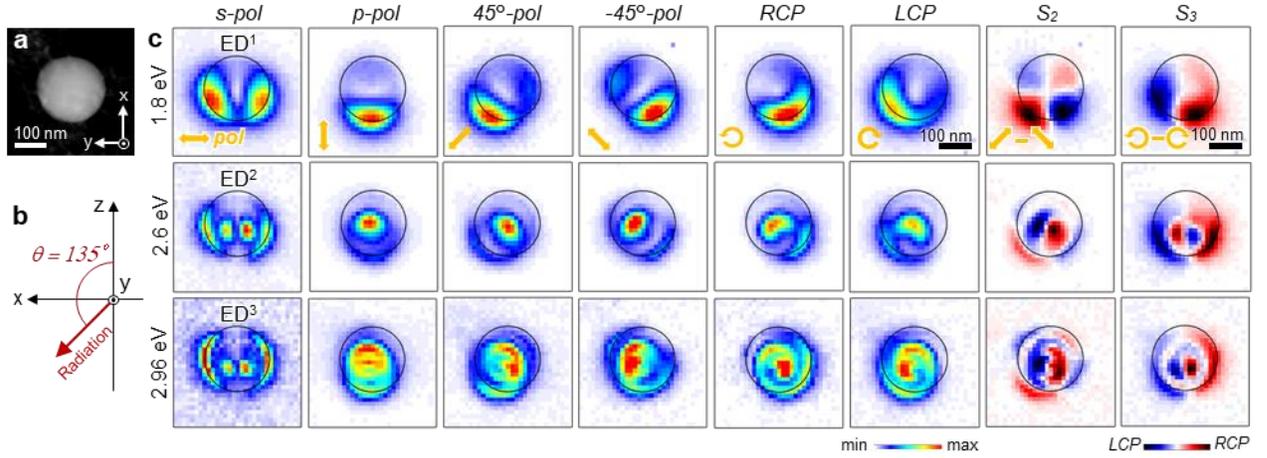

**Figure 3. CL maps of interfering *i*ED and *p*ED modes.** (a) STEM dark-field image of the probed Si nanosphere with a diameter of 250 nm (same particle as in Figure 2c). (b) Schematic illustration of the detection configuration. The CL signal is integrated over photon energy within a ±0.1 eV interval at a detection angle $\theta = 135°$. (c) Photon maps obtained for selected polarization angles $\zeta = 90°$ (*s*-pol), $0°$ (*p*-pol), $45°$ (RCP), and $-45°$ (LCP), along with the Stokes parameter $S_2$ and $S_3$ at photon energies of 1.8 eV (ED$^1$), 2.6 eV (ED$^2$), and 2.96 eV (ED$^3$). The edge of the Si nanosphere is indicated by black circles. The direction of polarization is schematically shown by yellow arrows in each panel. The emitted light is propagating normal to the image from the back side (see coordinate axes in (a) and (b)).

Since CPL generation is based on off-phase interference of two orthogonal fields, it is essential to determine the relative phase of the involved orthogonal modes. From the fully polarimetric dataset shown in Figure 3, one can quantitatively extract the relative phase of the two orthogonally polarized fields. Before discussing experimental phase measurements, we perform an analytical calculation to show that the *p*-polarized light actually works as a *reference* for mapping the phase of the in-plane modes at the detection angle $\theta = 135°$. We follow an analytical multipole decomposition (AMD) approach based on previously described methods to calculate the CL emission from homogeneous spheres[21, 26] now extended to include electron trajectories penetrating the particle (see full details in SI). Our theory allows us to calculate the complex field emission amplitude decomposed in the contributions coming each and all of the individual electric and magnetic modes labeled by orbital numbers ($l$, $m$) and the radial order. In Figure 4a-c, we show calculated line profiles of the phase of the polarization-projected emission amplitude for an electron moving along the $z$ direction and its position scanned along the $y$ axis across the center of a 250 nm Si sphere. As in experiment, we fix the detection angle at $\theta = 135°$ (Figure 3b). For comparison, the space- and angle-integrated spectra of the $i$ED contribution and the contribution from all modes are shown in Figure 4e. Note that the calculated spectra are slightly blue-shifted compared to experiment, presumably due to slight differences in size or dielectric constant (we use literature values in our calculations[34]). The phase patterns for $s$-polarization (Figure 4a and 4b) exhibit phase inversions when $y$ moves from negative to positive values. For instance, around the $i$ED$^1$ resonance, the phase is constantly negative for $y < 0$ and positive for $y > 0$, indicating a specific selection of $i$ED modes when measuring light polarized along the $y$ axis ($y$-$i$ED). Around the energies of ED$^2$ and ED$^3$ modes, the sign of the phase is flipped multiple times along the radial direction, thus evidencing the corresponding nodes of each mode. Even when including of all the electric and magnetic modes (Figure 4b), the phase pattern shows dominant contributions from projected $y$-$i$ED modes around their resonances, thus yielding similar features as in Figure 4a. In contrast, the phase of the total field resolved when measuring $p$-polarization is almost constant over energy and space, as shown in Figure 4b (see more details in SI). This constant phase can work as a reference for phase measurements. We thus conclude that the relative phase measurement of $\delta$ should correspond to the phase of the $s$-polarization light amplitude for in-plane modes. Indeed, the calculated relative phase pattern in Figure 4d corresponds well with the features of the *s*-polarized signal phase.

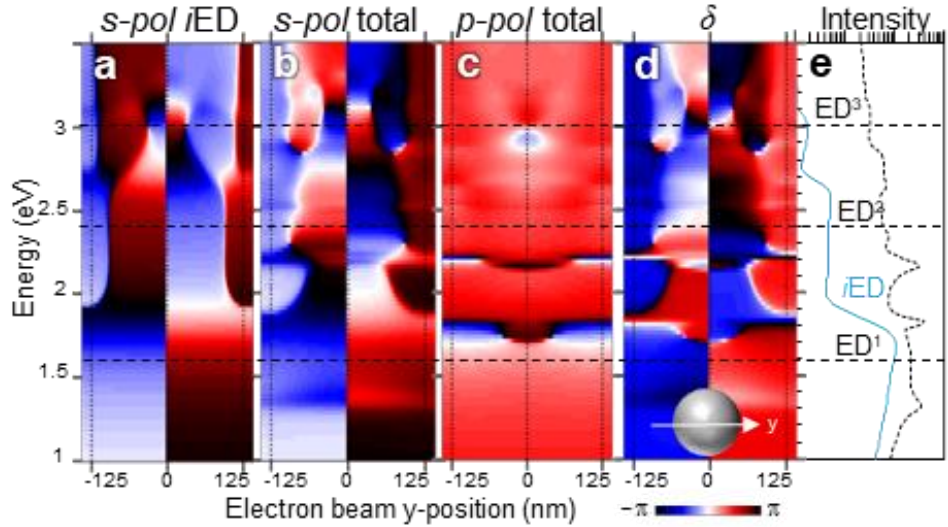

**Figure 4. Spectral line profile of the emission phase calculated from AMD theory.** (a-c) Calculated line profiles of the radiation phase of (a) the electric field extracted only from *i*EDs modes with *s*-polarization, (b) the total electric field with *s*-polarization (in-plane mode), (c) the total electric field with *p*-polarization, and (d) the relative phase $\delta$. The electron beam is scanned across the center of a 250 nm Si nanoparticle along the *y* direction. The detection angle is set to $\theta = 135°$. The black vertical dotted lines at *y* positions of -125 nm, 125 nm, and 0 nm indicate the edges and the center of the sphere. (e) Integrated spectra of all modes (dashed curve) and partial contribution of *i*EDs (light blue curve) plotted in semi-log scale. The energies of $ED^n$ modes, which correspond to the mapping in Figure 6, are indicated by horizontal dashed lines across all panels.

Based on the theoretical result that the phase of the *p*-polarized signal can be used as a nearly uniform reference, we experimentally perform relative phase mapping. Figure 5 shows the maps of the relative phase difference $\delta$, ellipticity angle $\eta$, and orientation angle $\alpha$ using the CL results of Figure 3b (see Methods section for details of the calculation procedure). For comparison, we also simulated the corresponding phase mapping by AMD calculation, which are shown in Figure 6. The experimental $\delta$ maps of Figure 5a show the relative phase difference of $y$-iEDs with respect to the $p$-polarized radiation, which is the reference at this detection angle, as we have discussed above. At the photon energy of $ED^1$ (1.8 eV), the sign of the phase is inverted with respect to the $x$-$z$ plane as shown schematically in Figure 5b. For $ED^1$, the phase distribution is not constant in the $x$ direction due to the influence of $i$ED modes projected along that axis ($x$-$i$ED), as demonstrated in the calculated phase of $p$-polarized components shown in Figure 6a. The calculated $\delta$ map in Figure 6 also shows the corresponding phase modulation along the $x$ axis, which reproduces well the experimental result. At photon energies of 2.6 eV ($ED^2$) and 2.96 eV ($ED^3$), the experimental $\delta$ maps (Figure 5a) of the higher-radial-order modes neatly show the phase flips between adjacent dipole *layers* corresponding to the poles of the $i$ED modes illustrated in Figure 5b. Also, for these higher ED modes, there is a slight phase modulation along the $x$ axis, which can be confirmed by the calculated phase maps in Figure 6b and 6c. For instance, at the energy of the $ED^3$ mode, the $\delta$ value changes from $-\pi/2$ (blue) through 0 (white) to $-\pi/2$ (blue), starting from the center to the $y$ positive direction (leftwards), without reaching the positive value of $\pi/2$ (red), unlike the perfect phase inversion illustrated in panel b (and vice versa for the right-hand side of red, black, and then red contrasts without reaching blue). This discrepancy from the *ideal* phase mapping illustrated in panel b is due to the intervention of higher-order multipoles ($l > 2$). The calculated relative phase $\delta$ reproduces the experimental result, including the aforementioned zero phase in the *second layer*, as shown in the line profile in Figure 4d and the $\delta$ map in Figure 6. Due to the influence of $x$-$i$EDs ($p$-pol), $\delta$ signals slightly deviate from those of pure $y$-$i$EDs ($s$-pol). In addition to the investigation of the phase difference $\delta$, fully polarimetric mapping can show the experimental distribution of the ellipticity angle $\eta$ and rotation angle $\alpha$. The resulting ellipticity maps can reveal the presence of CPL in the emission. Upon inspection of the $\eta$ maps, we conclude that almost full CPL ($\eta = \pm\pi/4$) can be achieved at the antinodes. We can thus attribute the generation CPL light to ED modes thanks to our photon mapping and relative-phase-extraction analyses.

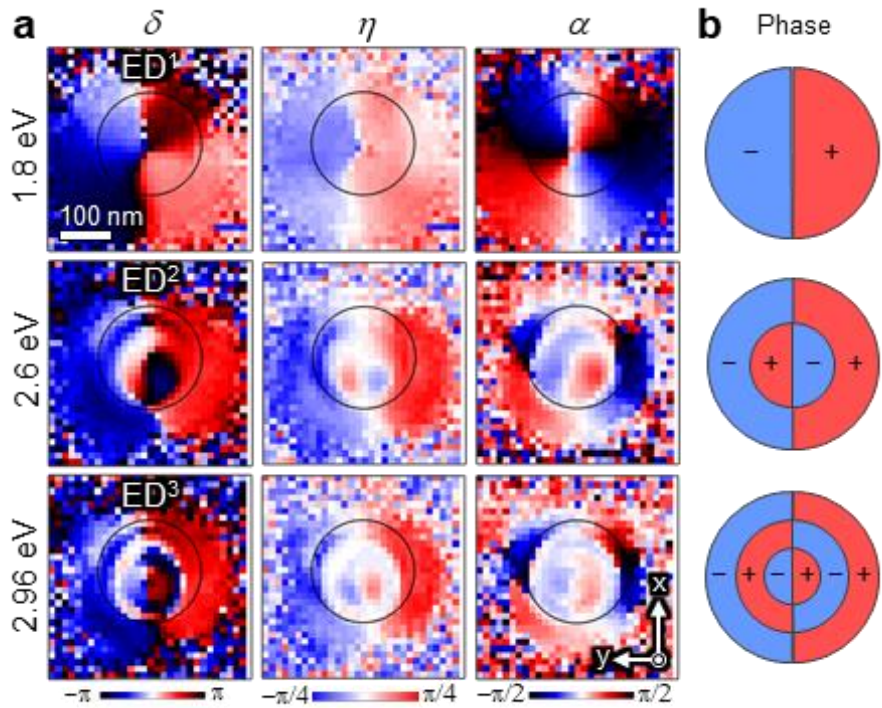

**Figure 5. Phase mapping of ED modes in the same Si sphere as in Figure 3.** (a) Maps showing the relative phase $\delta$, the ellipticity $\eta$, and the optical rotation $\alpha$ at a detection angle $\theta = 135°$ for photon energies of 1.8 eV (ED$^1$), 2.6 eV (ED$^2$), and 3.0 eV (ED$^3$), respectively. The phase $\delta$, obtained from the corresponding photon maps of Figure 3, is referred to the $p$-polarization component. (b) Schematic illustration of the phase sign in the corresponding electric dipole modes.
14

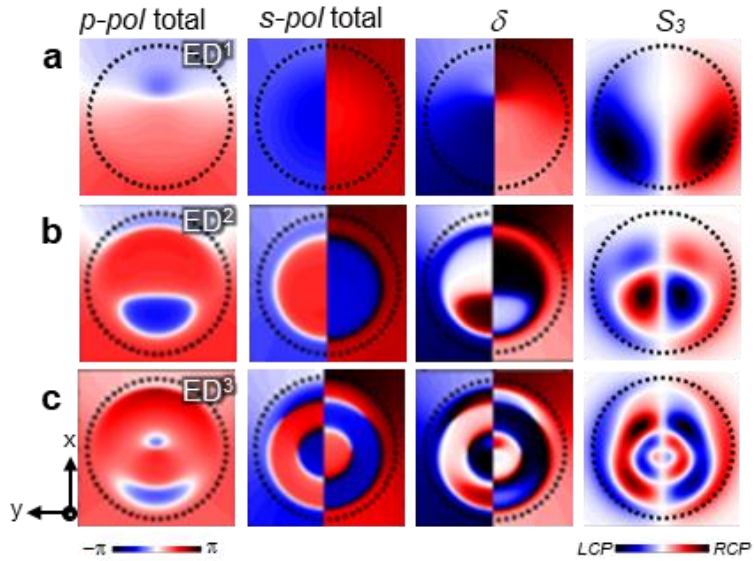

**Figure 6. Simulated phase maps of a 250 nm Si nanosphere with a detection angle $\theta = 135°$.** Panels a-c correspond to $ED^1$, $ED^2$, and $ED^3$ modes at photon energies of 1.6 eV, 2.4 eV, and 3.0 eV, respectively. The first and second columns from the left show the phase maps for *p*- and *s*-polarizations including all modes. The third and fourth columns display $\delta$ (the phase difference between *s*- and *p*-polarization CL emission components) and the Stokes parameter $S_3$. The edge of the sphere is indicated by a dashed circle in each panel.

**Angular Spectral Patterns of CPL Emission**

We have so far discussed the spatial distribution of CPL generation and the relative phase depending on the electron beam position. Since the angle of detection relative to the electron beam is another symmetry-breaking factor for CPL generation, the analysis of the angle-resolved emission amplitude including its phase can be important to gain control over CPL generation. We here investigate the angular dependence of CPL generation, which we can simultaneously perform with 2D spatial mapping of the same particle (Figure 3-5) using the 4D STEM-CL method. The angular distribution maps of the Stokes parameter $S_3$ and the phase difference $\delta$ are extracted for detection angles in the $\theta = 0 - 180°$ range at four representative positions along the $y$ direction across the center of the sphere (Figure 7). Gray masks overlay unobservable regions caused by specimen-shadowing ($\theta = 90°$) and the holes in the parabolic mirror around $\theta = 0$ and $180°$ through which the electron beam passes. For comparison, we show calculated angular spectral patterns using AMD theory. The sampling positions (i)-(iv) correspond to hotspots of $i$ED$^2$ modes. The phase difference $\delta$, shown in Figure 7a, varies depending on emission angle, photon energy, and excitation location in (i)-(iv); this dependence is also neatly reproduced by the simulation presented in Figure 7b. The experimentally measured $\delta$ values are somewhat noisy around $\theta = 0 - 90°$ at higher energies (> 2.0 eV) for edge-grazing excitations (i) and (iv) because the radiation amplitude is weak. This can be understood from the patterns $S_3$ (panels c and d), which include the actual CPL amplitude with $S_3 = 2I_p I_s \sin \delta$. Indeed, in the $S_3$ radiation patterns (Figure 7c) at the positions (i) and (iv), the CPL emission is stronger in the downward direction, except at low emission energies (< 2.0 eV). Interestingly, the observed parity is mostly the same for all the directions and photon energies at the edge. In contrast, the CPL strength and parity with excitation at the inner positions (ii) and (iii) (Figure 7c and 7d) dramatically vary depending on the photon emission angle and energy; at photon energies in the 2.0 - 2.6 eV range, the chiral emission has strong intensity depletions, while at higher photon energies of 2.6 - 3.0 eV strong CPL emission is observed with opposite parities. These trends are neatly reproduced by the AMD theory, as shown in Figure 7d. In a more detailed analysis presented in the SI, the directionality of the CPL radiation is found to be dominantly caused by the contribution of the EQ ($l, m$) = (2, 0) mode. Thus, these results demonstrate tunable CPL generation from a sphere by adjusting the detection angle, the electron beam position, and the photon energy.

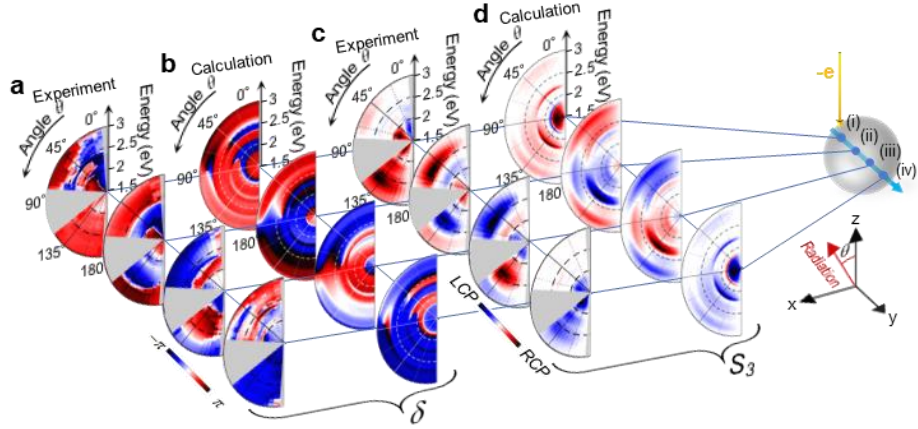

**Figure 7. Angular dispersion patterns of chiral emission from the same sphere as in Figure 3.** We show the experimentally measured (a,c) and calculated (b,d) phase differences $\delta$ (a,b) and $S_3$ (c,d) at beam-spot positions (i) $y = -120$ nm, (ii) $y = -45$ nm, (iii) $y = 45$ nm, and (iv) $y = 120$ nm along a scan following the $y$ axis and crossing the center of the sphere ($x = 0$ nm) as shown in the right inset. Angle ranges inaccessible in our setup are shaded in gray.

## CPL Generation by Interfering ED and MD Modes

In this section, we demonstrate CPL generation by relying on the concept of interfering MD and ED modes, as schematically illustrated in Figure 1b and 1d. This CPL generation scheme can be discerned by the CPL distribution with quadrupolar symmetry around the $z$ axis, which we experimentally visualize here. We choose a slightly smaller particle (170 nm diameter, see Figure 8a) than in the previous discussion in order to have the full profile of the low energy $MD^1$ mode within the measurable range (see Figure 8c). When observing a rotationally symmetric structure around the $z$ axis, such as the present sphere, rotating the detection angle $\varphi$ with the excitation electron beam position at a certain radial distance $r$ from the $z$ axis is equivalent to rotating the electron beam with that radius $r$ around the $z$ axis for fixed detection angle $\varphi$ (see details in SI).

Therefore, the information contained in the angular radiation pattern as a function of $\varphi$ can be obtained by analyzing the photon maps. Figure 8 shows photon maps of the Stokes parameter, the phase, and the polarization state parameters observed at a polar detection angle of $\theta = 10°$ for a measured photon energy of 2.18 eV, which lies between the MD and ED resonances (see Figure 8c). The detection angle $\varphi$ is fixed to 0°. We choose the polar angle $\theta = 10°$ because radiation at $\theta = 0°$ has no chirality, as described in Figure 1b and 1d (*i.e.*, the electric fields of $i\text{MD}^1$ and $i\text{ED}^1$ modes are parallel), and also because the $p\text{ED}^1$ modes is more strongly excited as the angle $\theta$ approaches 90°. The $S_3$ plot in Figure 8d exhibits four hotspots at the sphere edge with mirrored parities against the *x-y* and *y-z* planes. Now we extract the $\varphi$ directional angular plots by scanning the excitation beam position along a circle of radius $r = 78$ nm around the center of the sphere. We find that grazing electron beam excitation (*e.g.*, near the edge of the sphere) is efficient to excite the *i*ED and *i*MD modes dominantly. The obtained polar plot of $S_3$ is shown in Figure 8e. The quadrupolar symmetric pattern is clearly discernable for polar angles $\theta < 30°$, as anticipated in Figure 1d. We note that the CPL radiation of the first scheme (Figure 1a and c) becomes dominant at $\theta > 30°$ because the *p*ED mode is then more strongly excited. This quadrupolar symmetric pattern of CPL generation clearly demonstrates interference of magnetic and electric modes. Finally, we also note that this second strategy with interfering ED and MD modes should work with excitation by linearly polarized plane waves.

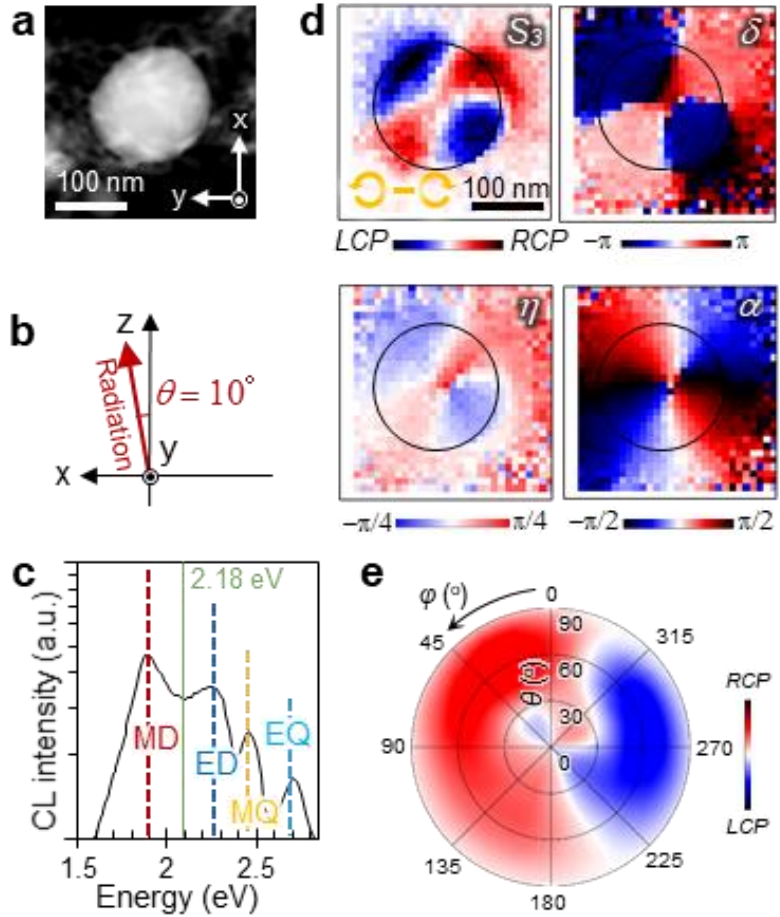

**Figure 8. CL maps of interfering ED and MD modes in a 170 nm Si particle.** (a) STEM dark-field image of the observed Si nanosphere. (b) Schematic illustration of the detection configuration. The CL signals are integrated over a photon energy interval of ±0.1 eV at the detection angle $\theta = 10°$. (c) CL spectrum without polarization selection, as obtained by integrating the signal over the radiation angular emission range $\theta = 0 - 180°$ and averaging over all electron beam positions on the sample. Mode types are indicated by labels in the spectral peaks. (d) Mapping of the Stokes parameter $S_3$, the phase difference $\delta$, the ellipticity $\eta$, and the orientation angle $\alpha$. The edge of the Si nanosphere is indicated by black circles on each image. (e) $S_3$ angular plot as a function of $\theta$ and $\varphi$, obtained by rotating the electron beam position around and at a radial distance $r = 78$ nm with respect to the sphere center for a measured photon energy of 2.18 eV.

**CONCLUSIONS**

The emission of CPL from a dielectric sphere has been realized by following two complementary generation mechanisms, which we have visualized using a fully polarimetric 4D STEM-CL setup. The first strategy consists in interfering degenerate in- and out-of-plane electric dipole modes with different phases controlled by the position of the electron beam. We measured full polarimetry maps and further extracted emission phase maps, clearly visualizing CPL generation from the Si sphere up to full circular polarization ($\eta = \pm\pi/4$), depending on electron beam positions. This mapping also revealed the distribution of emitted-light phase associated with the excitation of high-radial-order electric dipole modes even inside the sphere. A rigorous theoretical analysis based on a multipolar decomposition of the emitted optical fields supports this phase mapping approach and shows results in excellent agreement with experiment. The angular and spectroscopic distribution of CPL emission showed that the CPL parity can be fully controlled not only by the excitation position but also through the photon emission direction and energy. We further experimentally and theoretically demonstrated a second concept of CPL generation consisting in interfering electric dipole and magnetic dipole modes, showing characteristic quadrupolar symmetric in the angular profiles of the CPL emission. The use of 4D STEM-CL holds great potential for the development of customizable CPL sources, whereby the phase and degree of polarization of the light emission are readily controllable through positioning the electron beam on the sphere. Further customization is expected to be achievable by using other sample morphologies. The versatility of the present approach to generate light with controlled phase and polarization could find application in encoding information on the phase and polarization of emitted photons. Such an electron beam-based device could possibly be realized using the technologies developed for field emission displays.

## METHODS

### Detail of 4D STEM-CL

A modified STEM (JEM-2100F, JEOL, Japan) with a Cs-corrector is used at an acceleration voltage of 80 kV.[35] The electron probe current is about 1 nA with 20 mrad illumination half-angle, which results in a 1 nm probe size. These beam conditions are optimized so that we obtain sufficient CL signals without noticeable damage on the samples. An aluminum parabolic mirror is mounted between the pole pieces of the objective lens, so that the sample is exactly on the focal position of the mirror. (see Figure 2a) The radiation passing through the vacuum window is polarized by a polarization system consisting of a quarter-wave plate (QWP) and a linear polarizer to resolve circular polarization. The QWP is aligned such that the fast axis is tilted by 45° from the $z$ axis., where RCP is converted into horizontal polarization. A slit mask resolves the emitted light in the $\varphi$ direction while enabling us to simultaneously collect all angular information in the $\theta$ direction. Angular information along the $\theta$ direction is dispersed along the vertical axis on the 2D CCD surface by inserting a cylindrical lens in the optical system, which allows focusing the collimated beam in the $\varphi$ direction while imaging the $\theta$-selected intensity along the vertical axis on the CCD plane. The grating of the spectrometer disperses the photon energy in the horizontal direction of the CCD plane. The chromatic aberration induced by the optical system is corrected by post-processing the CCD image (see details in SI). The photon maps are extracted from the corrected angle- and energy-dispersion patterns.

**Normalization of Mapping Intensities for Phase Calculations**

Before calculating the Stokes parameters and the relative phases, we corrected the position drift and normalized the intensity of the image at each polarization. The following normalization procedure compensates the intensity unbalance in the 4D STEM setup caused by the optical elements and the asymmetry of the spectrometer. The position of each photon map is adjusted based on the STEM images of the particles acquired simultaneously with the photon maps. The normalization is performed on the basis of two conditions: i) The integrated intensity over all the mapping area is equal to that of the geometrically symmetric measurement with respect to the *x-z* plane, namely $\iint I_{45°}ds = \iint I_{-45°}ds$ and $\iint I_{RCP}ds = \iint I_{LCP}ds$; ii) The sum of the integrated intensity of orthogonal polarizations should be equal to the unpolarized intensity, that is, $\iint I_s ds + \iint I_p ds = \iint I_{45°}ds + \iint I_{-45°}ds = \iint I_{RCP}ds + \iint I_{LCP}ds$. These conditions lead to simpler normalization relations of the integrated intensities for ±45° and circular polarizations as $\iint I_{45°}ds = \iint I_{-45°}ds = \iint I_{RCP}ds = \iint I_{LCP}ds$. The phase shift caused by light reflection at the parabolic mirror is subtracted from the measured phase (see details in SI).

**Fabrication of Si Nanospheres**

Si nanospheres are fabricated by femtosecond laser ablation.[27] A Si wafer is ablated by a commercial 1 kHz femtosecond laser (Hurricane, Spectra Physics) with 1 mJ, 100 fs pulses at a central wavelength of 800 nm. The laser beam is focused down to 10 μm in diameter. The particles produced by laser irradiation are collected on a TEM grid with an elastic carbon membrane (~25 nm thickness), which is placed above the ablated Si wafer. To a good approximation, particles acquire a spherical shape due to surface tension during the fabrication process.

**Analytical Multipole Decomposition (AMD) Theory**

We extend a previously developed formalism for CL emission from a sphere excited by an external electron beam in order to include penetrating trajectories. Here, we summarize the resulting expressions, which are derived in detail in the SI for an electron moving with constant velocity $v$ along a straight-line trajectory parallel to and separated a distance $R_0$ from the $z$ axis, and crossing the $x$-$y$ plane at a point that forms an azimuthal angle $\varphi_0$ with respect to the $x$ axis. The sphere of radius $a$ and local, frequency-dependent permittivity $\epsilon(\omega)$ is taken to be centered at the origin of coordinates. This formalism allows us to write the electric far-field of the emitted light at the position $\mathbf{r}$ and time $t$ as $\mathbf{E}(\mathbf{r},t) = (1/r)\int(d\omega/2\pi)\ e^{i\omega r/c - i\omega t}\mathbf{f}_{\text{CL}}(\Omega,\omega)$, where $c$ is the speed of light in vacuum, $\Omega$ is the direction of emission, and

$$\mathbf{f}_{\text{CL}}(\Omega,\omega) = \frac{1}{k}\sum_L \left[\vec{\zeta}_L(\Omega)\psi_L^{\text{M,ind}} + \hat{\mathbf{r}} \times \vec{\zeta}_L(\Omega)\psi_L^{\text{E,ind}}\right] \quad (1)$$

is the angle- and frequency-dependent emission amplitude, decomposed in multipolar modes that are labeled by angular momentum numbers $L = (l, m)$ with symmetry $\nu = $ E (electric) and $\nu = $ M (magnetic). Also, $k = \omega/c$ is the free-space light wave vector. In this expression, we use vector spherical harmonics

$$\vec{\zeta}_L(\Omega) = \frac{1}{2}\left[C_L^- Y_{l,m-1}(\Omega) + C_L^+ Y_{l,m+1}(\Omega)\right]\hat{\mathbf{x}} + \frac{i}{2}\left[C_L^- Y_{l,m-1}(\Omega) - C_L^+ Y_{l,m+1}(\Omega)\right]\hat{\mathbf{y}} + m Y_L(\Omega)\hat{\mathbf{z}}$$

defined in terms of spherical harmonics $Y_L(\Omega)$ with $C_L^\pm = \sqrt{(l \pm m + 1)(l \mp m)}$. The scalar coefficients in eq 1 are given by (see detailed derivation in the SI)

$$\psi_L^{M,\text{ind}} = -e^{-im\varphi_0} \frac{4\pi i^{1-l} mek}{l(l+1)c} \left\{ t_l^M \frac{v}{c} A_L^+ K_m\left(\frac{\omega R_0}{v\gamma}\right) \right.$$

$$\left. + \int_{-z_0}^{z_0} dz\, e^{i\omega z/v} Y_L(\theta, 0)\left[-k\, t_l^M h_l^{(+)}(kr) + k' B_l^M j_l(k'r) - k\, j_l(kr)\right] \right\}$$

$$\psi_L^{E,\text{ind}} = -e^{-im\varphi_0} \frac{2\pi i^{1-l} ek}{l(l+1)c} \left\{ \frac{1}{\gamma} t_l^E B_L K_m\left(\frac{\omega R_0}{v\gamma}\right) \right.$$

$$+ \frac{1}{R_0} \int_{-z_0}^{z_0} dz\, e^{i\omega z/v} \left[ C_L^-(R_0 \partial_{R_0} + 1 - m)\left(-t_l^E h_l^{(+)}(kr) + \frac{B_l^E}{\sqrt{\epsilon}} j_l(k'r)\right.\right.$$

$$\left. - j_l(kr) \right) Y_{l,m-1}(\theta, 0)$$

$$\left.\left. - C_L^+(R_0 \partial_{R_0} + 1 + m)\left(-t_l^E h_l^{(+)}(kr) + \frac{B_l^E}{\sqrt{\epsilon}} j_l(k'r) - j_l(kr)\right) Y_{l,m+1}(\theta, 0) \right]\right\}$$

in which $r = \sqrt{R_0^2 + z^2}$, $\theta = \cos^{-1}(z/r)$, $j_l$ and $h_l^{(+)}$ are spherical Bessel and Hankel functions, respectively, and we use the Mie scattering and transmission coefficients

$$t_l^M = \frac{-\rho_1 j_l(\rho_0) j_l'(\rho_1) + \rho_0 j_l(\rho_1) j_l'(\rho_0)}{\rho_1 h_l^{(+)}(\rho_0) j_l'(\rho_1) - \rho_0 j_l(\rho_1)\left[h_l^{(+)}(\rho_0)\right]'}$$

$$t_l^E = \frac{-j_l(\rho_0)[j_l(\rho_1) + \rho_1 j_l'(\rho_1)] + \epsilon j_l(\rho_1)[j_l(\rho_0) + \rho_0 j_l'(\rho_0)]}{h_l^{(+)}(\rho_0)[j_l(\rho_1) + \rho_1 j_l'(\rho_1)] - \epsilon j_l(\rho_1)\left\{h_l^{(+)}(\rho_0) + \rho_0\left[h_l^{(+)}(\rho_0)\right]'\right\}}$$

$$B_l^M = -\rho_1 \frac{h_l^{(+)}(\rho_1) j_l'(\rho_1) - j_l(\rho_1)\left[h_l^{(+)}(\rho_1)\right]'}{\rho_0 j_l(\rho_1)\left[h_l^{(+)}(\rho_0)\right]' - \rho_1 h_l^{(+)}(\rho_0) j_l'(\rho_1)}$$

$$B_l^E = -\epsilon \rho_1 \frac{h_l^{(+)}(\rho_1) j_l'(\rho_1) - j_l(\rho_1)\left[h_l^{(+)}(\rho_1)\right]'}{\epsilon j_l(\rho_1)\left\{h_l^{(+)}(\rho_0) + \rho_0\left[h_l^{(+)}(\rho_0)\right]'\right\} - h_l^{(+)}(\rho_0)[j_l(\rho_1) + \rho_1 j_l'(\rho_1)]}$$

where $\rho_0 = ka$ and $\rho_1 = ka\sqrt{\epsilon}$. We further define the electron multipolar coupling coefficients

$$A_L^+ = i^{l+m}(2m-1)!! \sqrt{\frac{(2l+1)(l-m)!}{\pi(l+m)!}} \frac{(c/v)^m}{v\gamma^m} C_{l-m}^{m+1/2}\left[\frac{c}{v}\right],$$

$$B_L = C_L^+ A_{l,m+1}^+ - C_L^- A_{l,m-1}^+$$

where $C_{l-m}^{m+1/2}$ are Gegenbauer polynomials evaluated at the argument value $c/v$.[36, 37] The CL photon emission probability is readily obtained from the far-field Poynting vector divided by the photon energy $\hbar\omega$,[21, 26] which leads to

$$\Gamma_{CL}(\Omega,\omega) = \frac{1}{4\pi^2 \hbar k} |\mathbf{f}_{CL}(\Omega,\omega)|^2$$

This quantity is normalized in such a way that the integral over angle and frequency $\int d\Omega \int_0^\infty d\omega \, \Gamma_{CL}(\Omega,\omega)$ yields the total photon emission probability. Finally, using the orthogonality of the vector spherical harmonics (see SI), we obtain the angle-integrated CL photon emission probability

$$\Gamma_{CL}(\omega) = \int d\Omega \Gamma_{CL}(\Omega,\omega) = \frac{1}{4\pi^2 \hbar k^3} \sum_{L,\nu} l(l+1) |\psi_L^{\nu,\text{ind}}|^2$$

where the sum runs over electric ($\nu = E$) and magnetic ($\nu = M$) multipoles $L = (l, m)$.

**ASSOCIATED CONTENT**

**Supporting Information**

The Supporting Information is available free of charge at https://pubs.acs.org. Data analysis of angle- and energy-resolved patterns, phase shift caused by the aluminum parabolic mirror, 4D dataset of CL maps, analytical multipole decomposition (AMD) method, comparison of AMD calculations of CL spectra and Mie calculations of the optical scattering cross section, simulated phase line profiles, equivalence of beam rotation and detection rotation, angular plots revealing interference of ED and MD modes (PDF)

AUTHOR INFORMATION

Corresponding Author

Takumi Sannomiya − Department of Materials Science and Technology, Tokyo Institute of Technology, 4259 Nagatsuta Midoriku, Yokohama 226-8503, Japan; orcid.org/0000-0001-7079-2937; Email: sannomiya.t.aa@m.titech.ac.jp

Author Contributions

The manuscript was written through contributions from all authors. TM and TS conceived the research, carried out the experiment, and performed the analysis. FJGA developed the AMD theory and computational tool. All authors have given approval to the final version of the manuscript.

ACKNOWLEDGMENTS

We thank Dr. Taka-aki Yano, Dr. Masaki Hada and Dr. Takuo Tanaka for helping us with the sample fabrication, Mr. Jaewon Shin for improving the CCD image acquisition software, and Dr. Naoki Yamamoto for his valuable advices. The authors acknowledge the financial support from JST PRESTO (#JPMJPR17P8), JSPS PD2(#20J14821), ERC (advanced grant 789104-eNANO), the Spanish MINECO (MAT2017-88492-R and SEV2015-0522), the Catalan CERCA, and Fundació Privada Cellex. TM is grateful for the JSPS Research Fellowship for Young Scientists.